\documentstyle[aps]{revtex}
\begin{document}
%\draft  % Makes pacs numbers print (REVTEX)
%
\title{Phase Diagram of a Model of Correlated Hopping of Electrons in a
Lattice of Berry Molecules}
\vspace{10mm}
\author{Giuseppe Santoro,$^{1,2}$ Nicola Manini,$^{1,2}$
Alberto Parola,$^{1,3}$ and Erio Tosatti$^{1,2,4}$}
\address{$^1$ Istituto Nazionale di Fisica della Materia (INFM)}
\address{$^2$ International School for Advanced Studies (SISSA), Via Beirut 4,
I-34013 Trieste, Italy}
\address{$^3$ Istituto di Fisica, Via Lucini 3, Como, Italy}
\address{$^4$ International Centre for Theoretical Physics (ICTP), P.O. Box
586, I-34014 Trieste, Italy}
%%%%%%%%%%%%%%%%%%%%%%%%%%%%%%%%%%%%%%%%%%%%%%%%%%%%%%%%%%%%%%%%%%%%%%%
%
%\date{\today}
\maketitle
\vspace{10mm}
\begin{abstract}
The $1D$ phase diagram of a model for correlated hopping of electrons in a
lattice of Berry phase molecules is presented.  Electrons hop in presence
of an extra orbital degree of freedom at each site. This is mimicked as a
spin-1 variable whose allowed states depend on the electron occupancy so as
to take into account the orbital degeneracies of different molecular
occupancies.  In the $1D$ case we find that at low electron densities $n<<1$
there is a region with dominant superconducting correlations surviving an
additional repulsive on-site interaction $U$ as strong as the bandwidth,
$W=4t$.  The critical value $U_c$ of $U$ below which superconductivity is
found to be dominant decreases with increasing density $n$. For $n=1/2$ we
find $U_c/t\approx 1$, whereas at $n=1$ (half-filling) our (less accurate)
results are compatible with $U_c/t\approx 0$.  For $U>U_c(n)$ and away from
half-filling ($n\neq 1$) the system is metallic with dominant $2k_F$ charge
density wave (CDW) correlations.  At half-filling a charge gap opens for
$U>U_c$ and the system becomes an insulator.  A spin-gap characterizes the
phase-diagram for all densities and for all values of $U$, even in the
metallic regime $U>U_c$.
\end{abstract}
\pacs{PACS numbers: 71.27.+a,74.20.Mn,74.70.Wz}
%
%%%%%%%%%%%%%%%%%%%%%%%%%%%%%%%%%%%%%%%%%%%%%%%%%%%%%%%%%%%%%%%%%%%%%%%%%%
% TESTO
%%%%%%%%%%%%%%%%%%%%%%%%%%%%%%%%%%%%%%%%%%%%%%%%%%%%%%%%%%%%%%%%%%%%%%%%%%
\newpage

\section{Introduction}

In the normal theory of metals not much attention is usually devoted to
orbital phenomena. The atomic orbital angular momentum is first of all
largely quenched by crystal fields, which split part of the orbital
degeneracy. Subsequently, the electron kinetic energy, usually large,
completes the job by eliminating residual degeneracies except at
special points in the Brillouin zone.

The scope of this paper is to explore and describe in some detail a
possible scenario emerging in the opposite limit. This is realized
considering ``molecular'' sites with a well-defined orbital degeneracy,
unquenched by crystal fields, and with a coherent, but extremely weak,
electron hopping energy $t$ between the sites.
The specific extra ingredients which we consider are: (a) the orbital
degeneracy depends on electron occupancy; (b) inter-site electron hopping
forces a switch of orbital degeneracy for both the initial and final site.

This kind of model is closely inspired by our previous
studies\cite{AMT,attachment} of the physics of C$_{60}^{n-}$ ions, present
in molecular fullerides. An isolated C$_{60}^{n-}$ ion undergoes a
dynamical Jahn-Teller (JT) effect, which, as we found, is affected by an
electronic Berry phase when $n$ is odd, but not when it is even. As a
result, the molecular ground state has $L=1$ in the former case and $L=0$
in the latter as in point (a) above.

A realistic model for the superconducting compounds A$_3$C$_{60}$ (A=K,Rb)
should take into account the electronic and molecular degrees of freedom of
C$_{60}$: the three degenerate $t_{1g}$ electronic orbitals (LUMO) at each
molecule and the eight $H_g$ 5-dimensional phonon multiplets which are
allowed to couple, by symmetry, with the electronic orbitals.  The energies
of these phonons range from $30$ to $200$ meV, i.e., comparable to the
typical electronic bandwidth $W\approx 0.4$ eV in the solid.  In such a
situation, a standard Eliashberg approach, heavily based on Migdal's
theorem (neglect of electron-phonon vertex corrections) and on the
smallness of $\hbar\omega_{\rm Debye}/W$, is hard to justify from first
principles,\cite{Pietronero,Gunnarsson} even if empirically
successful.\cite{Ramirez}

We explore here a different route, based on the opposite limit in which the
hopping matrix element $t$ is the smallest energy scale in the problem.
The basic ingredients of the model we will study are the different
degeneracies associated to molecular states with different electronic
occupation, as suggested by a careful treatment of the JT effect
of each molecule.\cite{AMT,Gunnarsson,MTD} While well aware of the
limitations of such an approach -- the real situations being in the
intermediate regime $t\approx \hbar\omega$ -- we still believe it
illustrates the potentially interesting role played by orbital degeneracies
in molecular crystals.

The basic ideas can be illustrated in a model which is much simpler than
that of C$_{60}$, i.e., a system of ``$e\otimes E$'' molecules\cite{englman}
with two electronic orbitals per site $c_{\pm,\sigma}$, coupled to a
two-dimensional JT active phonon multiplet $b_{\pm}$ of energy
$\hbar\omega$:
\begin{eqnarray} \label{original_model:eqn}
H &=& -t\,\sum_{\pm} \sum_{<r,r'>} \sum_{\sigma}
\, [\, c^{\dagger}_{\pm,\sigma}(r') c_{\pm,\sigma}(r) + H.c. \, ] \,+\,
\hbar \omega \sum_{\pm} \sum_{r}
\, [\, b^{\dagger}_{\pm}(r) b_{\pm}(r) + \frac{1}{2}\, ] \nonumber \\
&& \hspace{20mm} +\; H_{\rm e-ph} \;+\; (\mbox{e-e terms}) \nonumber \\
H_{\rm e-ph} &=& \frac{g \hbar \omega}{2} \sum_{r} \sum_{\sigma}
\{ \, c^{\dagger}_{+,\sigma}(r) c_{-,\sigma}(r) \,
[ b^{\dagger}_{-}(r) + b_{+}(r) ]  \,+\, H.c. \} \;.
\end{eqnarray}
Such a simplified model has been considered in Ref.\ \cite{MTD} and studied
using a strong coupling ($g\rightarrow \infty$) approach to the
electron-phonon molecular Hamiltonian.  Consider first the case in which
there is no hopping ($t=0$).  The molecular Hamiltonian commutes with an
angular momentum-like operator
\begin{equation}
j(r) \,=\, [ \, b^{\dagger}_{+}(r) b_{+}(r) - b^{\dagger}_{-}(r) b_{-}(r) \, ]
\;+\; \frac{1}{2}\sum_{\sigma}
[ \, c^{\dagger}_{+,\sigma}(r) c_{+,\sigma}(r) -
     c^{\dagger}_{-,\sigma}(r) c_{-,\sigma}(r) \, ] \;.
\end{equation}
One finds that molecules with an {\em odd\/} number of electrons $n$
($n=1,3$) have an associated electronic Berry phase of $\pi$ and a
four-fold degenerate ground state ($j=\pm 1/2,\sigma=\pm 1/2$).  On the
contrary, molecules with an even number of electrons ($n=0,2$) have no
Berry phase and a non-degenerate singlet ground state ($j=0,S=0$).  For
$g>>1$, excited molecular states are separated from the ground state by
terms of order $\hbar\Omega = \hbar\omega/g^2$.\cite{englman}

However, the validity of this analysis is not restricted to a strong
coupling regime.  For $g<<1$, conventional $2^{nd}$-order perturbation
theory in $H_{\rm e-ph}$, neglecting retardation effects, gives an
effective phonon-mediated electron-electron interaction of the type
\cite{Tsvelik}
\begin{equation} \label{eff_inter:eqn}
H_{\rm e-e} = -\frac{g^2\hbar\omega}{4} \sum_{r} \sum_{\sigma,\sigma'} \;
\{ \, c^{\dagger}_{+,\sigma}(r) c_{-,\sigma}(r)
      c^{\dagger}_{-,\sigma'}(r) c_{+,\sigma'}(r) \;+\; H.c. \} \;.
\end{equation}
It is simple to verify that the four degenerate $n=1$ states
$c^{\dagger}_{\pm,\sigma} |0\rangle$ have an energy $E_{n=1}\approx
-g^2\hbar\omega/4$, whereas the lowest $n=2$ state is the inter-orbital
singlet $(c^{\dagger}_{+,\uparrow} c^{\dagger}_{-,\downarrow} -
c^{\dagger}_{+,\downarrow} c^{\dagger}_{-,\uparrow}) |0\rangle$ with energy
$E_{n=2}\approx -g^2\hbar\omega$.\cite{Higher_n=2:nota}
Once again, ($j=\pm 1/2,\sigma=\pm 1/2$) for the $n=1$ states, whereas
($j=0,S=0$) for the lowest $n=2$ state.

Equation \ref{eff_inter:eqn} shows that the inclusion of the JT
effect proves already important in providing a sizable {\em pairing
energy\/} which acts as a kind of effective negative Hubbard-$U$ term,
therefore reducing the effect of the local Hubbard repulsion between
electrons.\cite{AMT} It is therefore not surprising that, {\em in absence
of strong repulsive e-e interactions\/}, this phonon-mediated local
attraction leads to superconductivity. Manini, Tosatti and Doniach have
found that this pairing attraction is important at least as long as
$t<\hbar\Omega$.  Shelton and Tsvelik have studied, using bosonization in
$1D$, the purely electronic model consisting of the hopping term $T$ and
the effective non-retarded electron-electron interaction $H_{\rm e-e}$ in
Eq.\ \ref{eff_inter:eqn}.\cite{Tsvelik} They too find a strong tendency to
superconductivity. However, these negative Hubbard $U$'s can in practice be
easily canceled with Coulomb repulsive Hubbard $U$'s.

We will show here that there are effects, induced by orbital degeneracy,
which can lead to superconductivity even when {\em repulsive interactions
overwin the polaronic attraction\/}. These effects will be found to be more
pronounced at low carried density.

As argued above, singly occupied states have an extra degeneracy of orbital
origin ($j=\pm 1/2$) while the lowest doubly occupied state is
non-degenerate ($j=0$). If the hopping $t$ is much smaller than the lowest
local excitation energy, we can effectively restrict the Hilbert space at
each site by keeping only the ground state for each electronic
occupation.\cite{MTD} We are thus led to considering a Hubbard-type model
in which each site of the lattice carries an extra pseudospin-1 degree of
freedom, with the following constraint: a singly-occupied site has, in
addition to the usual spin degeneracy, a twofold (orbital) degeneracy
represented by $S^z=\pm 1$, whereas a doubly occupied (or empty) site must
have $S^z=0$ and is therefore non-degenerate,
\begin{eqnarray} \label{CONSTRAINT:eqn}
n_r=1\; \; &\longrightarrow& \; S^z_r=\pm 1 \;\; , \nonumber \\
n_r=0,2 \; &\longrightarrow& \; S^z_r=0 \;.
\end{eqnarray}
The electron-pseudospin (EP) working Hamiltonian $H$ is written
as follows:
\begin{equation} \label{MODEL:eqn}
H \,=\, - \frac{t}{2} \sum_{<rr'>}\sum_{\sigma}
\; (c^{\dagger}_{r\sigma} c_{r'\sigma} + H.c.) \, (S^+_r S^-_{r'} + H.c.)
\,+\, U \sum_r n_{r\uparrow} n_{r\downarrow} \;,
\end{equation}
where the $S^{\pm}_r$'s are spin-1 ladder operators at each site, and the
remaining notation is completely standard.
It is worth stressing that while $H$, and in particular the hopping term,
{\em conserves the constraint\/} in Eq.\ \ref{CONSTRAINT:eqn}, the model
is still highly non-trivial even for $U=0$.

% Michele Fabrizio. Bound states all'ordine g^4 ... missing at g^2

Manini, Tosatti and Doniach have introduced and studied this EP model in
one-dimension and at half-filling \cite{MTD}.  By small ring exact
diagonalizations and a BCS-like mean field calculation they show that the
effect of the correlated hopping, even in the absence of any negative-$U$
term, is to induce pairing between electrons of opposite spin.

In this paper we present the results we have obtained for the full phase
diagram of the model in $1D$, which is illustrated in Fig.\ 1.  We find
that at low densities $n<<1$ there is a region with dominant
superconducting correlations surviving repulsive interactions as strong as
the bandwidth, $W=4t$.  The critical value $U_c$ of $U$ below which
superconductivity is found to be dominant decreases with increasing density
$n$. For $n=1/2$ we find $U_c/t\approx 1$, whereas at half-filling our
results (not very accurate, due to marginal umklapp terms) are compatible
with $U_c/t\approx 0$.  For $U>U_c(n)$ and out of half-filling ($n\neq 1$)
the system is metallic with dominant $2k_F$ charge density wave (CDW)
correlations.  At half-filling a charge gap opens up for $U>U_c$ and the
system becomes an insulator.  A spin-gap characterizes the phase-diagram
for all densities and for all values of $U$, even in the metallic regime
$U>U_c$.

\section{Some definitions and a few exact results, mainly in $1D$}

As we will discuss more in detail in Section 3, only charge degrees of
freedom remain gapless in our model. As a consequence, the large-distance
physics distances is characterized by a quantity $K_{\rho}$ -- determining
the behavior of the correlation functions decaying as power-laws -- and by
the velocity $u_{\rho}$ of the sound-like collective
excitations.\cite{Hal_81,Krho_notation}
As we will see in Section \ref{Phased:sect}, the relevant correlation functions
are density-density ($N(x)$) and singlet superconductive ($S(x)$),
behaving as
\begin{eqnarray} \label{correlations_1:eqn}
N(x) &\,=\,& -\frac{K_{\rho}}{2\pi^2} \frac{1}{x^2} +
A_2 \frac{ \cos{(2k_F x)} } { x^{K_{\rho}} }
+ A_4 \frac{ \cos{(4k_F x)} }{ x^{4K_{\rho}} } \,+\,  \cdots \nonumber \\
S(x) &\,=\,& \frac{B_2}{x^{1/K_{\rho}}} \,+\, \cdots \;.
\end{eqnarray}

For general values of $U$ and of the density $n$, results for
$K_{\rho}$ can be obtained in a standard way \cite{Schulz}
from a knowledge of the charge velocity $u_{\rho}$,
\begin{equation} \label{urho_1:eqn}
u_{\rho} \,=\, \lim_{L\rightarrow \infty} \, \frac{L}{2\pi} \,
[E_L(S=0,k=2\pi/L)-E_L] \;,
\end{equation}
of the inverse compressibility $\kappa^{-1}$
\begin{equation} \label{compressibility_1:eqn}
\frac{1}{\kappa n^2} = \lim_{L\rightarrow \infty}
L\frac{\partial^2 E_L}{\partial N_c^2} =
\frac{\pi}{2} \frac{u_{\rho}}{K_{\rho}} \;,
\end{equation}
and of the Drude peak strength $D_c$ \cite{Schulz,Shastry}
\begin{equation}\label{drude_1:eqn}
D_c \,=\, \lim_{L\rightarrow \infty}
\left. \frac{L}{2} \frac{d^2 E_L(\Phi)}{d\Phi^2}\right|_{\Phi=0}
\,=\, \frac{1}{\pi} u_{\rho} K_{\rho} \;.
\end{equation}
Here $E_L(S=0,k=2\pi/L)$ is the energy of the lowest excited singlet state
with total momentum $k=2\pi/L$ with respect to the ground state of energy
$E_L$ in a ring of size $L$, and $E_L(\Phi)$ is the ground state energy
in presence of a magnetic flux $\Phi$.

In general, we resort to exact diagonalizations of the model on small rings
of size $L$ (with $L$ up to $16$), and use Eq.\ \ref{urho_1:eqn} to
estimate $u_{\rho}$ and Eq.\ \ref{compressibility_1:eqn} or
\ref{drude_1:eqn} to extract $K_{\rho}$. \cite{Krho:comment}
Three regions of the phase diagram are, however, well under control
using different approaches: the low density limit $n\rightarrow
0$,\cite{low_dens} and the limit of large repulsion or attraction
$U\to\pm\infty$.\cite{large_u,Hal_80} We summarize here, for the reader' s
convenience, the most important results concerning these three regions.

\subsection{The low-density limit: solution for two electrons}
\label{low_dens:sect}

Consider the two-electron problem first.  A state in the two-particle
Hilbert space with total z-component of the spin $M^z_{TOT}=0$ (for both
the electrons and the spin-1 states) can be written as
\begin{equation}
|\Psi\rangle \,=\, \sum_{r,r'} \;
[ \, \psi_{+-}(r,r') \, S^{+}_r S^{-}_{r'} \,+\,
     \psi_{-+}(r,r') \, S^{-}_r S^{+}_{r'} \, ] \;
c^{\dagger}_{r\uparrow} \, c^{\dagger}_{r'\downarrow} \; |0\rangle \;,
\end{equation}
where the vacuum $|0\rangle$ is the state without fermions and with $S^z_r=0$
at each site.
In writing $|\Psi\rangle$ one takes into account the two possibilities of
associating a $S^z=\pm 1$ spin state to the up and down electrons:
$\psi_{+-}$ is the amplitude for having $S^z=+1$ associated to the
$\uparrow$-electron (and $S^z=-1$ to the $\downarrow$-electron), while
$\psi_{-+}$ is the amplitude for other possible choice.
The Schr\"odinger equation for $\psi_{+-}(r,r')$ is
\begin{eqnarray} \label{Sch:eqn}
E \psi_{+-}(r,r') &\,=\,& -t\, \sum_{a}
\, [ \, \psi_{+-}(r+a,r') + \psi_{+-}(r,r'+a) \, ]
\,+\, U \, \delta_{r,r'} \, \psi_{+-}(r,r') \nonumber \\
&& \hspace{04mm}
-t\, (\sum_{a} \delta_{r+a,r'}) \;
[ \, \psi_{-+}(r,r) + \psi_{-+}(r',r') \, ] \;,
\end{eqnarray}
where $a$ denotes a nearest neighbor vector ($a=\pm 1$ in $1D$).
A similar equation is obtained for $\psi_{-+}(r,r')$ by just exchanging
$\psi_{-+}$ and $\psi_{+-}$ everywhere.
The last term in Eq.\ \ref{Sch:eqn} is the crucial outcome of the extra
(orbital) degeneracy of singly occupied states, and deserves a few
comments.  When the two electrons are far enough in the otherwise empty
lattice, the EP Hamiltonian $H$ simply allows the hopping to a nearest
neighbor site of the ``composite'' object (see Fig.\ 2a) formed by an
electron and the associated spin-1 state (first term in Eq.\
\ref{Sch:eqn}). In other words, each electron retains its orbital index
during the hopping process.  Things are more subtle when two electrons come
to the same site $r$.  In such a case (see Fig.\ 2b), from a doubly
occupied site with $S^z_r=0$ one can reach, upon hopping, two possible
final states: either each electron keeps its own spin-1 label or the spin-1
labels associated to the two electrons are exchanged.
It is precisely this second possibility of exchanging spin-1 states
that is responsible for the presence of $\psi_{-+}$ in the equation for
$\psi_{+-}$ and vice-versa (last term in Eq.\ \ref{Sch:eqn}).

The Schr\"odinger equation is easily solved (on a square lattice in any D),
in momentum space, where it reads:
\begin{equation}
[E-2\epsilon_k] \, \psi_{+-}(k) \,=\, \frac{U}{L^D} \sum_p \psi_{+-}(p) +
\frac{2\epsilon_k}{L^D} \sum_p \psi_{-+}(p) \;.
\end{equation}
Here the case of total momentum $P=0$ has been considered, for simplicity,
and $\epsilon_k$ is the tight-binding dispersion of the free-electron problem
($\epsilon_k=-2t\sum_{\alpha}\cos{k_{\alpha}}$).
Introducing the quantity $J_{+-}=L^{-D}\sum_p \psi_{+-}(p)$, and the
corresponding one for the $-+$ amplitude, and imposing self-consistency, we
arrive at a simple $2\times 2$ problem
\begin{equation}
J_{+-} \,=\, J_{+-} \, \frac{U}{L^D} \sum_k \frac{1}{E-2\epsilon_k}
\,+\, J_{-+} \, \frac{1}{L^D} \sum_k \frac{2\epsilon_k}{E-2\epsilon_k} \;.
\end{equation}
(The equation for $J_{-+}$ is obtained by exchanging $J_{+-}$ with $J_{-+}$
everywhere.)
The set of solutions among which the ground state is found have
$J_{+-}=J_{-+}$ and their eigenvalues $E$ satisfy the equation
\begin{equation} \label{E:eqn}
\frac{1}{L^D} \sum_{k} \frac{1}{E-2\epsilon_k} = \frac{2}{E+U} \;.
\end{equation}
It is worth mentioning that in the ordinary Hubbard case, the right hand
side of Eq.\ \ref{E:eqn} would simply read $1/U$ \cite{creta}.  A graphical
analysis of Eq.\ \ref{E:eqn} readily shows that a bound state solution is
present even for $U>0$ up to $U_c=4Dt$ in $D\leq 2$.  In $D\geq 3$ a finite
attractive $U$ is needed to produce a bound state.

The bound state solution has been worked out analytically in the $1D$
case.\cite{low_dens} The bound state wavefunction naturally provides a
picture of bound pairs approximately localized on adjacent lattice
sites.\cite{low_dens} Remarkably, the rather strong attraction responsible
for this binding is {\sl generated by the kinetic term alone} via the
presence of the additional degrees of freedom.  The critical value of the
Hubbard repulsion [$U_c=4t$ in $1D$] is considerably larger than the ground
state binding energy at $U=0$ [$E_b/t=(8/\sqrt{3}-4)\sim 0.618$ in $1D$]
showing that this kind of pairing mechanism is rather insensitive to the
presence of on-site Coulomb repulsion.  The same feature is also present in
two dimensions where $U_c=8t$: The enhancement is due to the the larger
coordination of the 2D lattice which provides an even more efficient
delocalization of the electron pair.

This interpretation of the two-particle ground state, plus the additional
finding that there is no phase separation,\cite{low_dens} leads to a simple
picture, at least for $U<U_c$, of the low density limit of model
(\ref{MODEL:eqn}) both in one and two dimensions: the system behaves as a
weakly interacting, dilute gas of bosons with an extended core.  A
superfluid ground state must be expected at zero temperature.  In one
dimension, of course, off-diagonal long range order cannot occur and only a
long range power-law decay of the density matrix is possible, while in two
dimensions a genuine $T=0$ Bose condensate will form.  In terms of the
original electrons this implies a standard strong coupling BCS
superconducting ground state with localized Cooper pairs.  A similar
scenario has also been proposed in the framework of the one dimensional
$t-J$ model where bound pairs are formed at low density \cite{zurigo} and
$2<J/t<2.95$. In that case, however, the model is unstable to phase
separation, which in fact occurs massively at larger values of $J/t$,
whereas the present model shows no such tendency at least in
$1D$.\cite{low_dens}

{}From a low-density expansion of the ground-state energy and of the charge
velocity an analytic determination of the exponent $K_{\rho}$ in the limit
$n\rightarrow 0$ has been given in Ref.\ \cite{low_dens}.  It is found that
$K_{\rho}=2$ for $U<U_c(n=0)=4t$, whereas $K_{\rho}=1/2$ for
$U>4t$.\cite{Krho_notation}

\subsection{$U\to +\infty$ limit} \label{uinf:sect}

The strong coupling expansion of the EP model has
been worked out in Ref.\ \cite{large_u}.
The resulting Hamiltonian, to $O(t^2/U)$, is defined in the subspace of
empty or singly occupied sites where electrons
are characterized by {\it two} internal spin-$1/2$ degrees of freedom:
The usual spin ($\sigma$) and a pseudospin ($\tau$).
The latter takes into account the two possible states of the original
spin--1 rotators allowed by the constraint (\ref{CONSTRAINT:eqn}).
The resulting Hamiltonian, exactly like in the $t-J$ model, contains
a kinetic term (with coupling constant $-t$) and a spin (and pseudospin)
dependent contribution which scales as $t^2/U$.
In $1D$, analogously to the Hubbard model case, \cite{Ogata}
the total wavefunction factorizes, at $U\to\infty$, for all densities:
The position of the electrons is determined by a (free) spinless fermion
wavefunction while the spin (and pseudospin) ordering on the squeezed chain
is governed by the effective spin Hamiltonian
\begin{equation} \label{spinham}
H_{\sigma\tau}= - J_{\rm eff} \sum_{<i,j>}
\left [ 2 \vec\sigma_i\cdot \vec\sigma_j -{1\over 2} \right ]
\left [ 2 \vec \tau_i\cdot \vec \tau_j -{1\over 2} \right ]
\end{equation}
where the effective coupling constant depends on the electron density $n$
as $J_{\rm eff} = {2 t^2\over U} n [1-\sin(2\pi n)/2\pi n]$.  It turns out
that, in contrast with the Heisenberg case, the Hamiltonian in Eq.\
\ref{spinham} has a (doubly degenerate) valence bond ground state and is
characterized by a spin gap of order $J_{\rm eff}$.\cite{large_u}
Therefore, we conclude that our model (Eq.\ \ref{MODEL:eqn}), at arbitrary
(finite) density has a spin gap for $U\to\infty$.  The charge degrees of
freedom are, on the contrary, {\em gapless\/} and tend to approach the
spinless fermion case as $U\to\infty$.

Quantities such as the charge velocity and Drude peak
can be obtained analytically for any lattice size $L$ by this large-$U$
mapping of the charge degrees of freedom to free spinless fermions. The
expression for the charge velocity $u_{\rho}$ reads:
\begin{equation} \label{urho_inf}
u_{\rho}(U=+\infty) \,=\, 2t \frac{L}{\pi} \sin(\pi/L) \sin(\pi n)
\,\stackrel{L\rightarrow \infty}{\longrightarrow}\, 2t \sin(\pi n) \;,
\end{equation}
whereas the Drude peak strength $D_c$ is given by
\begin{equation} \label{drude_inf}
D_c(U=+\infty) \,=\, t \frac{ \sin(\pi n) }{ L \sin(\pi/L) }
\,\stackrel{L\rightarrow \infty}{\longrightarrow}\,\frac{t}{\pi} \sin(\pi n)
\;.
\end{equation}
As a result, $K_{\rho}=1/2$ for all densities $n\neq 1$ on the boundary line
$U=+\infty$.\cite{Krho_notation}

\subsection{$U\to -\infty$ limit} \label{uminf:sect}

The third, rather trivial, region where the physics is under control is
that of large on-site attraction.  Indeed, in the limit $U\rightarrow
-\infty$, only doubly occupied and empty sites are allowed. The extra
degeneracy of singly occupied sites plays therefore no special role, and
the model behaves as the corresponding negative $U$ Hubbard
model.\cite{UNEG:nota} In analyzing our data in this limit we will use, for
comparison, the numerical results for the Hubbard case obtained from Bethe
Ansatz.\cite{Schulz} At half-filling, the mapping of the $U\rightarrow
-\infty$ limit into the antiferromagnetic Heisenberg chain (with
$J=\frac{8t^2}{|U|}$ in the present case\cite{UNEG:nota}) provides
analytic $L=\infty$ results for the Drude peak $D_c$ and the charge
velocity $u_{\rho}$.\cite{Hal_80,Shastry} We get \cite{Drude:nota}
\begin{eqnarray} \label{uminf_results}
u_{\rho} &=& \frac{\pi}{2} \, J +\cdots \,=\,
\frac{\pi}{2} \frac{8t^2}{|U|} \,+ \, \cdots \nonumber\\
D_c &=& 4\, \frac{J}{8} +\cdots \,=\, \frac{1}{2} \frac{8t^2}{|U|}+\cdots \;.
\end{eqnarray}
(The dots indicate higher order corrections in $t/U$.)
The corresponding value of $K_{\rho}$ at half-filling for large negative
$U$ is therefore given by $K_{\rho}=1$.\cite{Krho_notation}

\section{The Phase Diagram ($1D$)} \label{Phased:sect}

In order to study the model using standard $1D$ Fermi gas techniques, we
have to represent the extra spin--1 degeneracy and the associated
constraint in a suitable way. A possible procedure is to first represent
the Hamiltonian in Eq.\ \ref{MODEL:eqn} in terms of additional hard-core
boson operators $d_{r\sigma}$ as follows:
\begin{equation} \label{MODEL_cd}
H \,=\, - t \sum_{<rr'>}\sum_{\sigma,\sigma'}
\; \{ \, c^{\dagger}_{r\sigma}  c_{r'\sigma}
         d^{\dagger}_{r\sigma'} d_{r'\sigma'} + H.c. \, \}
\,+\, \frac{U}{2} \sum_r [\, n^{c}_{r\uparrow} n^{c}_{r\downarrow} \,+\,
n^{d}_{r\uparrow} n^{d}_{r\downarrow} \,] \;,
\end{equation}
the constraint now being that, at each site,
\begin{equation} \label{CONSTRAINT_cd}
n^{d}_{r} \,=\, n^{c}_{r} \;.
\end{equation}
Eqs.\ \ref{MODEL_cd} and \ref{CONSTRAINT_cd} give a faithful representation
of all the matrix elements of the original model. The idea behind this
representation is that the extra label $S^z=+1$ (or $-1$) at each
singly-occupied site $r$ is represented by a hard-core boson
$d^{\dagger}_{r\uparrow}$ (or $d^{\dagger}_{r\downarrow}$). Empty sites
have no bosons, whereas the $S^z=0$ state of a doubly occupied site is
represented by two bosons
$d^{\dagger}_{r\uparrow}d^{\dagger}_{r\downarrow}$.

In $1D$, we can transform the hard-core bosons $d_{r\sigma}$ into spinful
fermions, by a Wigner-Jordan transformation.\cite{Shastry}
We are then led to consider a problem of correlated hopping of two species of
fermions in $1D$ with a conserved constraint (Eq.\ \ref{CONSTRAINT_cd}).
A possible way of treating this problem is to enlarge the Hilbert space by
removing the constraint in Eq.\ \ref{CONSTRAINT_cd} and to introduce a term
$V\sum_r ( n^{c}_{r} - n^{d}_{r} )^2$ with $V\to \infty$ in the Hamiltonian.
The correlated hopping term can then be regarded as a second-order process
obtained from a standard free hopping of the $c$ and $d$ particles as
follows:
\begin{eqnarray} \label{MODEL_ccd:eqn}
H_V &\,=\,& - \tilde{t} \sum_{<rr'>} \sum_{\sigma}
\, \{ \, c^{\dagger}_{r\sigma} c_{r'\sigma} \,+\,
         d^{\dagger}_{r\sigma} d_{r'\sigma} + H.c. \, \}
\,+\, \frac{U}{2} \sum_r [\, n^{c}_{r\uparrow} n^{c}_{r\downarrow} \,+\,
n^{d}_{r\uparrow} n^{d}_{r\downarrow} \,] \nonumber \\
&& \hspace{20mm}
\,+\, V \sum_r (\, n^{c}_{r} - n^{d}_{r} \,)^2 \,+\, \Delta H_V \nonumber \\
\Delta H_V &\,=\,& \frac{\tilde{t}^2}{V} \,
\sum_{<rr'>}\sum_{\sigma,\sigma'} \,
                          \{ \, c^{\dagger}_{r'\sigma} c_{r\sigma}
                                c^{\dagger}_{r\sigma'} c_{r'\sigma'} \,+\,
                                d^{\dagger}_{r'\sigma} d_{r\sigma}
                                d^{\dagger}_{r\sigma'} d_{r'\sigma'}  \, \} \;,
\end{eqnarray}
where $\tilde{t}^2/V=t$, and $V$ is understood to be large.  The extra
term $\Delta H_V$ is included in order to cancel unwanted terms generated
by second-order perturbation theory in the hopping $\tilde{t}$.  We have
thus obtained a reasonably standard form of two-chain problem, which we
treat in weak-coupling in $U$ and $V$ (assuming continuity with the large $V$
regime), by linearizing the band around the
two Fermi points at $\pm k_F$ and introducing right ($p=R=+$) and left
($p=L=-$) fermion fields $\psi_{pa\sigma}(x)$ for every species of fermions
($a=c,d$, and $\sigma=\uparrow,\downarrow$). Luckily, the ``g-ology''
\cite{Solyom} of this two-chain problem turns out to be a particular case
of the two-chains in presence of transverse hopping treated by Fabrizio
{\em et al.\/}.\cite{Fabrizio} We can therefore directly use the RG
equations of Ref.\ \cite{Fabrizio} calculated up to third order in the
couplings (two-loops). Away from half-filling (i.e., when umklapp terms are
irrelevant) the result is that the model scales to a strong-coupling fixed
point and gaps open in all sectors with the exception of the totally
symmetric one, which remains gapless.
Correlation functions can be calculated by bosonizing the strong coupling
fixed point.\cite{Fabrizio} We introduce a bosonic representation of the
fermion fields
\begin{equation}
\psi_{pa\sigma}(x) \,=\,
\frac{ e^{ipk_Fx} }{(2\pi\alpha)^{1/2}} \,
e^{ i\sqrt{\pi} [\Theta_{a\sigma}(x)+p\Phi_{a\sigma}(x)] }\;,
\end{equation}
where $\Phi_{a\sigma}(x)$ and $\Theta_{a\sigma}(x)$ are boson fields such
that $\Pi_{a\sigma}(x)=-\nabla \Theta_{a\sigma}(x)$ is the momentum
conjugate to $\Phi_{a\sigma}(x)$, and $\alpha$ is a short-distance cut-off.
In terms of charge/spin and symmetric/antisymmetric combinations of the
fields
\begin{eqnarray}
\Phi_{a\rho(\sigma)} &\,=\,& \frac{1}{\sqrt{2}}
[ \Phi_{a\uparrow} \pm \Phi_{a\downarrow} ] \nonumber \\
\Phi^{\pm}_{\rho(\sigma)} &\,=\,& \frac{1}{\sqrt{2}}
[ \Phi_{c\rho(\sigma)} \pm \Phi_{d\rho(\sigma)} ] \;,
\end{eqnarray}
the bosonized Hamiltonian for the gapless field $\Phi^+_{\rho}$ reads
\begin{equation} \label{boson_h:eqn}
{\cal H} \,=\, \frac{1}{2} \,
\int_0^L \! dx \, [\, u_{\rho} K^+_{\rho} (\Pi^+_{\rho})^2
+ \frac{u_{\rho}}{K^+_{\rho}} (\nabla \Phi^+_{\rho})^2 \} \;+\;
\frac{\pi}{8L} [\, \frac{u_{\rho}}{K^+_{\rho}} (\hat{N}-N_o)^2 +
		   u_{\rho} K^+_{\rho} \hat{J}^2 \,] \;,
\end{equation}
whereas the other fields are effectively locked to their strong-coupling
values $\Phi^-_{\rho}=\Phi^{\pm}_{\sigma}=0$. Here $\hat{N}$ is the
operator counting the total number of particles of both species ($N_o$ is
its ground state value), and $\hat{J}$ is the associated current
operator.\cite{Hal_81} We do not calculate $u_{\rho}$ and $K^+_{\rho}$
perturbatively, since we are interested in the physical limit $V\to \infty$.
We will therefore extract the correct $u_{\rho}$ and $K^+_{\rho}$ numerically
from finite size exact diagonalizations (see below).

{}From Eq.\ \ref{boson_h:eqn} the compressibility relation can be
immediately derived.\cite{Hal_81,Schulz} It reads:
\begin{equation}
L\frac{\partial^2 E_L}{\partial N^2} \,=\,
\frac{\pi}{4} \frac{u_{\rho}}{K^+_{\rho}} \;,
\end{equation}
where $N=2N_c$ is the total number of particles of both species.
Similarly, the Drude peak is found to be given by
\begin{equation}
D_c \,=\, \frac{1}{2\pi} u_{\rho} K^+_{\rho} \;.
\end{equation}

It is simple to verify that the correlation functions which decay as a
power-law at large distances are the density-density correlations
$N(x)=\langle n^{c}_x n^{c}_0 \rangle$ and the singlet superconductive
correlations $S(x)=\langle P^{\dagger}_x P_0 \rangle$ with $P_x =
\psi_{c\uparrow}(x) \psi_{c\downarrow}(x) \psi_{d\uparrow}(x)
\psi_{d\downarrow}(x)$.
Expressing the fermionic fields in terms of the bosonic ones, and using the
Hamiltonian in Eq.\ \ref{boson_h:eqn}, we readily find that:
\begin{eqnarray} \label{correlations:eqn}
N(x) &\,=\,& -\frac{K^+_{\rho}}{4\pi^2} \frac{1}{x^2} +
A_2 \frac{ \cos{(2k_F x)} } { x^{K^+_{\rho}/2} }
+ A_4 \frac{ \cos{(4k_F x)} }{ x^{2K^+_{\rho}} } \,+\,  \cdots \nonumber \\
S(x) &\,=\,& \frac{B_2}{x^{2/K^+_{\rho}}} \,+\, \cdots \;.
\end{eqnarray}
We emphasize that all the relations obtained so far, from the compressibility
to the correlation functions, have exactly the same form they
would take in the negative-$U$ Hubbard case \cite{Solyom,upos:nota}
if we define $K_{\rho}=K^+_{\rho}/2$.
The corresponding forms of the correlation functions, compressibility, and
Drude peak are then given by Eqs.\
\ref{correlations_1:eqn},\ref{compressibility_1:eqn},\ref{drude_1:eqn}.
(In the compressibility relation, a factor $4$ is absorbed by the fact that
$N=2N_c$, $N_c$ being the number of electrons.)  To facilitate the
comparison with the Hubbard case we will use in the following the notation
$K_{\rho}=K^+_{\rho}/2$, so that the latter will be directly comparable to
the Hubbard model's $K_{\rho}$.\cite{upos:nota}

{}From Eq.\ \ref{correlations:eqn} we observe that regions with
$K_{\rho}(=K^+_{\rho}/2)>1$ and $K_{\rho}<1$ are characterized by dominant
correlations (i.e., correlations with the slowest decay) of the
superconductive and $2k_F$-CDW type, respectively.  We recall that, in the
low-density limit $n\to 0$, $K_{\rho}=\frac{1}{2}$ for $U>U_c=4t$ and
$K_{\rho}=2$ for $U<U_c$.  This defines the left ($n=0$) border of the
phase diagram in Fig.\ 1.  For $U<U_c$ the dominant correlations are
superconductive, decaying as $1/\sqrt{x}$, exactly as in a
dilute ($n\rightarrow 0$) hard core boson gas.  For $U>U_c$, the
correlation function with the slowest decay is the $2k_F$ density-density
response, with an inverse square root behavior $1/\sqrt{x}$, implying a
divergence in the $2k_F$-CDW susceptibility, except at $U=\infty$ where the
$2k_F$ response function has vanishing amplitude, like in the Hubbard
model, and the corresponding singularity is therefore absent.

\subsection{Numerical results}

We are now ready to discuss the results of the finite-size exact
diagonalization data of the EP Hamiltonian (\ref{MODEL:eqn}).
First we will show that a {\em spin gap\/} -- which is expected all over
the phase diagram, and not just in the superconducting region -- is indeed
found for all the values of $U$ studied, at $n=1/2$ and $n=1$.
Figs.\ \ref{Spingap1/2:fig} and \ref{Spingap1:fig} show the finite-size gaps
between the ground state and the lowest triplet state, $\Delta E_L(S=1)$, as
a function of the inverse lattice size $1/L$ for both the quarter-filled
(Fig. \ref{Spingap1/2:fig}) and the half-filled case (Fig. \ref{Spingap1:fig})
and for several values of $U$.
As expected, $\Delta E_L(S=1)$ always extrapolate to a finite non-zero value
in the infinite volume limit for all the values of $U$ studied.
The extrapolated values of the spin gap $\Delta E_{\infty}(S=1)$
(see Fig. \ref{Spingap1:fig} - inset for the $n=1$ case) decrease with
increasing $U$ in agreement with the result that, as $U\rightarrow +\infty$,
the spin gap should be proportional to
$J_{\rm eff}=\frac{2t^2}{U}[n-\sin{(2\pi n)}/(2\pi)]$.\cite{large_u}

Next we turn to the exponent $K_{\rho}$ for finite values of the density
$n$.  The exact eigenvalues of $L$-site rings provide determinations of the
three quantities $u_{\rho}$ (Eq.\ \ref{urho_1:eqn}), $D_c$ (Eq.\
\ref{drude_1:eqn}), and $\partial^2 E_{L}/\partial N_c^2$ (Eq.\
\ref{compressibility_1:eqn}).
In particular, the magnetic flux $\Phi$ through the ring in the evaluation
of the Drude peak (\ref{drude_1:eqn}) is implemented simply by changing the
hopping matrix element $t\to t e^{i\Phi/L}$ (a procedure equivalent to
twisting the boundary conditions) and calculating the second derivative
with respect to $\Phi$ numerically.  Moreover, as usual, the second
derivative of the energy with respect to the number of electrons $N_c$,
appearing in the compressibility (\ref{compressibility_1:eqn}), is
approximated by the finite difference expression
$L[E_L(N_c+2)+E_L(N_c-2)-2E_L(N_c)]/4$.

Figs.\ \ref{Drude1/2:fig} and \ref{Drude1:fig} show the finite-size values
of the Drude peak $D_c$ and of the charge velocity $u_{\rho}$ (inset) for
the quarter-filled ($n=1/2$) and half-filled ($n=1$) case, for several
values of $U$.
The $U=+\infty$ values of $D_c$ and
$u_{\rho}$ for $n=1/2$ (see Eqs.\ \ref{drude_inf} and \ref{urho_inf}) are
shown in Fig.\ \ref{Drude1/2:fig} by solid lines.
In the quarter-filling case -- representative of the $0\leq n<1$ region of
the phase diagram -- a finite non-zero Drude peak, signal of a metallic
ground state, is obtained for any $U$.  Correspondingly, the charge
velocity $u_{\rho}$ scales to a finite value for any $U$.
On the contrary, at half filling and for positive enough $U$ ($n$=1,
Fig.\ \ref{Drude1:fig}) the Drude peak vanishes in the thermodynamic
limit, indicating insulating behavior.
This is due to the opening of a gap in the charge
sector above some critical value of $U$ close to 0.
In Fig.\ \ref{Chargegap:fig} the finite-size charge gap at half-filling
\begin{equation} \label{charge_gap:eqn}
\Delta_c \,=\, \frac{1}{2} [ E(N_c=L+2) - E(N_c=L) - U ]
\end{equation}
is plotted: $\Delta_c$ scales nicely to 0 as $L\to \infty$ for negative $U$,
whereas it extrapolates towards finite nonzero
values (roughly proportional to $U$ for large $U$) for positive enough $U$.
The extrapolated
values of the charge gap for several values of $U$ are shown in the inset
of Fig. \ref{Chargegap:fig}, together with the positive $U$ Hubbard model
values, for comparison.
Finally, Figs.\ \ref{Krho1/2:fig} and \ref{Krho1:fig} show the finite-size
estimates of the exponent $K_{\rho}$\cite{Krho:comment} for $n=1/2$ and
$n=1$, respectively, and for several values of $U$.
For the quarter-filled case (Fig.\ \ref{Krho1/2:fig}) it is instructive to
compare $K_{\rho}$ with the corresponding Hubbard model result, obtained
from a numerical solution of the Bethe Ansatz equations.\cite{Schulz} The
arrows in panel (a) of Fig.\ \ref{Krho1/2:fig} indicate the values of
$K_{\rho}$ for the Hubbard model at the corresponding value of $U$.  The
$U/t=-100$ data confirm that the EP model behaves like the Hubbard model in
the $U\to -\infty$ limit.

The most important information, however, is contained in the $U=0$ data of
Fig.\ \ref{Krho1/2:fig}: the exponent $K_{\rho}$ at $U=0$ scales to a value
somewhat larger than the critical value 1, which is also the value of
$K_{\rho}$ for the non-interacting Hubbard model.  This indicates (see Eq.\
\ref{correlations:eqn}, with $K_{\rho}=K^+_{\rho}/2$) that our model has
{\em dominant superconducting correlations even in absence of attractive
interactions.\/} Panel (b) of Fig.\ \ref{Krho1/2:fig} shows that the
superconducting state survives, for $n=1/2$, up to a positive $U/t$ of
order 1.
Fig.\ \ref{Krho1:fig} shows the values of $K_{\rho}$ at half-filling,
together with the Hubbard model results (solid squares).  At half-filling,
umklapp processes are marginal even in the gapless regime, introducing
logarithmic corrections to the size scaling.\cite{Cardy} The data for
the Hubbard model show clearly the consequence of such a non-trivial size
scaling: an extrapolation based on the data at small sizes gives sizable
errors (of the order of 10\%).  In view of this non-trivial size scaling,
it is hard to make definite statements on the critical value of $U$ above
which a charge gap opens and the system starts to be an insulator.
Nevertheless, the $U=0$ data for $K_{\rho}$ (panel (b) of Fig.\
\ref{Krho1:fig}) seem to be definitely below the value of $1$, where
umklapp processes should become relevant.  We are therefore inclined to
believe that a charge gap should be present even at $U=0$ (see also the
comparison with the Hubbard results of Fig.\ \ref{Chargegap:fig}), although
the $U=0$ Drude peak results of Fig.\ \ref{Drude1:fig} might suggest a
conducting behavior.

The phase diagram in Fig.\ \ref{diagram:fig} summarizes our results on the
model in Eq.\ (\ref{MODEL:eqn}) in $D=1$. The most remarkable feature is in
the low density region $n\ll 1$, where superconducting correlations are
found to be dominant up to positive values of the on-site coulomb repulsion
$U$ comparable to the electronic bandwidth.  We believe that the physics at
small densities is robust.  The bound states discussed in Section
\ref{low_dens:sect} should lead to Bose condensation of strongly localizes
Cooper pairs in $2D$. In $3D$ no bound state is found but a superconducting
instability should still occur in the presence of a Fermi surface.  On the
other hand, other features of the model, like the presence of a spin gap
over all the phase diagram -- including the positive $U$ metallic regime --
are probably confined to the low dimensionality $1D$ case.

\section{Conclusions}

We have studied the $1D$ phase diagram of a model of electrons which hop
weakly between molecular sites. Due to the Berry phase mechanism, each
molecular site possesses an unquenched orbital degeneracy, only when its
electron occupancy is odd. Upon electron hopping, the orbital degeneracies
must correspondingly switch, as the occupancies of the initial and final
states change. By adding adjustable on-site Hubbard $U$ term, we can
effectively remove from the problem all polaronic effects, and we can thus
study the effect of orbital degeneracy, and of its hopping-induced switch,
in isolation.
The working model (\ref{MODEL:eqn}) is therefore Hubbard-like, but embodies
the orbital degeneracy part in the form of a pseudospin-1 on-site
variable. We have studied its zero-temperature phase diagram as a function
of electron filling and of Hubbard $U$.

Through a series of mappings -- all of which are exact for what concerns
the asymptotic behavior of correlations in $1D$ -- we find that the problem
is equivalent to a particular variant of two coupled $1D$ chains, whose
behavior is generally well studied\cite{Fabrizio}. This mapping predicts
that there will generally be no charge gap (except at half filling for
$U>U_c$) while there will always be a spin gap, for all $n>0$ and all
$U$. The zero charge gap is related to either sliding CDW conductivity or
superconductivity, and we have distinguished these two different regions in
the phase diagram.
A quantitative study of the phase diagram, conducted by exact diagonalization
and finite size scaling, has led to the overall picture of
Fig. \ref{diagram:fig}.

The most interesting fact in the phase diagram is the dominance of
superconductive behavior extending even into the (moderately)
repulsive-$U$ region. Pairing is associated with new orbital correlations
-- as foreshadowed by an exact 2-electron solution -- which permit a
lowering of kinetic energy through an enhanced nearest neighbor pair
scattering. This feature represents an important novelty brought about by
orbital degeneracy, and should reasonably carry over to more complex and
realistic situations in higher dimensions.
This pairing mechanism is not easily destroyed by a repulsive $U$, it is more
effective at low carrier density, and is apparently immune from the polaron
self-trapping, which depresses $T_c$ in strongly coupled electron-phonon
systems.

Another new feature found in this model in $1D$ is the presence of an
overall spin gap, including the nonsuperconductive region of the phase
diagram, where sliding CDW $2k_F$ correlations dominate. It is at this
stage not clear in what way this behavior (which appears rather specific of
1D) will change in higher dimensions.

ACKNOWLEDGEMENTS --
It is a pleasure to thank M. Fabrizio, S. Sorella, and S. Doniach for many
enlightening discussions. E. Tosatti acknowledges the sponsorship of NATO,
through CRG 920828.

%%%%%%%%%%%%%%%%%%%%%%%%%%%%%%%%%%%%%%%%%%%%%%%%%%%%%%%%%%%%%%%%%%%%%%%%%
%                               BIBLIOGRAPHY
%%%%%%%%%%%%%%%%%%%%%%%%%%%%%%%%%%%%%%%%%%%%%%%%%%%%%%%%%%%%%%%%%%%%%%%%%

%%%%%%%%%%%%%%%%%%%%%%%%%% FIGURES %%%%%%%%%%%%%%%%%%%%%%%%%%%%

\begin{figure}
\caption{The $1D$ phase diagram for the EP model
(Eq.\ \protect\ref{MODEL:eqn}) discussed in the present paper.  For details
see Sect. \protect\ref{Phased:sect}. }
\label{diagram:fig}
\end{figure}

\begin{figure}
\caption{The hopping mechanism of an up-electron starting from a singly (a)
or doubly (b) occupied site. }
\label{Hop:fig}
\end{figure}

\begin{figure}
\caption{The spin-gap, defined as the difference in energy between the
lowest triplet state and the ground state (singlet), for $n=1/2$ and
$U=0,2t$ and $4t$, as a function of the inverse lattice size $1/L$}
\label{Spingap1/2:fig}
\end{figure}

\begin{figure}
\caption{The spin-gap for $n=1$ and $U/t=0,1,4,10$ and 100. Inset: the
extrapolated values of the spin-gap in the thermodynamic limit. The
extrapolations are obtained through weighted quadratic polynomial fit to
finite size data. The asymptotic $1/U$ behavior is apparent at the right
side of the plot.}
\label{Spingap1:fig}
\end{figure}

\begin{figure}
\caption{The Drude peak $D_c$ and the
charge velocity $u_{\rho}$ (inset) for $n=1/2$ (quarter-filling).
The lines for the Drude peak are fit to the finite-size data.
The large-$U$ behavior (solid line) is discussed in
Sect. \protect\ref{uinf:sect}.}
\label{Drude1/2:fig}
\end{figure}

\begin{figure}
\caption{The Drude peak $D_c$ and the
charge velocity $u_{\rho}$ (inset) at half-filling ($n=1$). The data for
$U=2t$ and $10t$ show clear indication of $D_c$ vanishing in the thermodynamic
limit, while for negative $U$, $D_c$ remains finite.
The data for $U=0$ seem also to converge to a nonzero value, but there are
evidences (see Sect.\ \protect\ref{Phased:sect}) in favour of insulating
behavior ($D_c=0$).  }
\label{Drude1:fig}
\end{figure}

\begin{figure}
\caption{The charge gap at half filling, obtained as the difference in
energy between the lowest $N_c=L+2$ state and the $N_c=L$ ground state
(Eq.\ \protect\ref{charge_gap:eqn}).
The inset reports the extrapolations obtained through weighted
quadratic polynomial fit to finite-size data, together with the Hubbard model
values, for comparison.
The charge gap seems to vanish at some point close to $U=0$. }
\label{Chargegap:fig}
\end{figure}

\begin{figure}
\caption{The exponent $K_{\rho}$ for the EP model
at quarter filling obtained (Eq.\ \protect\ref{compressibility_1:eqn}) as
$K_{\rho}=\pi D_c / u_{\rho}$.  Panel (a): $U/t=-100,-4$ and 0; arrows mark
the asymptotic values of $K_{\rho}$ for the Hubbard model obtained by Bethe
Ansatz. Panel (b): $U=t$ and $4t$; squares mark the scaling of $K_{\rho}$ for
the Hubbard model at the corresponding values of $U$, for comparison.}
\label{Krho1/2:fig}
\end{figure}

\begin{figure}
\caption{The exponent $K_{\rho}$ for the EP model at half filling obtained
(Eqs.\ \protect\ref{compressibility_1:eqn} and \protect\ref{drude_1:eqn})
as $K_{\rho} = \pi \left( D_c / 2 L \frac{\partial^2 E_{GS}}{\partial
N_c^2} \right )^{1/2}$.  Square dots represent the scaling of the same
quantity for the Hubbard model at the corresponding values of $U$. Notice
the non-trivial size scaling, due to marginal umklapp terms.  Panel (a):
$U=-20t$ and $-2t$. Panel (b): $U=0$ and $t$.}
\label{Krho1:fig}
\end{figure}


\begin{thebibliography}{99}
\bibitem{Higher_n=2:nota}
Higher $n=2$ states are either intra-orbital singlets
$c^{\dagger}_{\pm,\uparrow} c^{\dagger}_{\pm,\downarrow} |0\rangle$,
with $E_S \approx -g^2\hbar\omega/2$, or triplet states ($E_T=0$).
\bibitem{Krho_notation}
The notation adopted here differs from that of Ref.\ \cite{low_dens} where
$K_{\rho}$ is used to indicate what we call here $K^+_{\rho}$. In this paper
we use $K_{\rho}$ to indicate the quantity $K^+_{\rho}/2$, which is
directly comparable to the Hubbard model's $K_{\rho}$.
\bibitem{Krho:comment}
Introducing the finite-size values of the three quantities $L\partial^2
E_{L}/\partial N_c^2$, $u_{\rho}$ and $D_c$ in relations
(\ref{compressibility_1:eqn}) and (\ref{drude_1:eqn}) we get, in general,
two independent estimates of the exponent $K_{\rho}$.
\bibitem{UNEG:nota}
For $U\rightarrow -\infty$, in particular, the model maps into a Heisenberg
problem with $J=8t^2/|U|$ instead of simply $4t^2/|U|$ as in the Hubbard case.
This is due to the additional degeneracy of the singly occupied states which
appear as intermediate states in second order perturbation theory.
\bibitem{Drude:nota}
The extra-factor $4$ in the expression for $D$, relative to the Heisenberg
case, is due to the fact that if a flux $\Phi$ is included in the hopping,
$t\rightarrow t e^{i\Phi/L}$, the $J$-coupling is modified as
$J\rightarrow J e^{2i\Phi/L}$ so that
$E_{U\rightarrow -\infty}(\Phi) = E_{\rm Heisenberg}(2\Phi) $, generating a
factor $4$ in the Drude peak, related to the second derivative of the
energy.
\bibitem{upos:nota}
For the positive-$U$ Hubbard case, due to the absence of a spin gap, the
density-density and superconductive correlation functions are not given
by Eq.\ \protect\ref{correlations_1:eqn} in terms of $K_{\rho}$.
The relationships determining $K_{\rho}$
(Eqs.\ \protect\ref{compressibility_1:eqn},\protect\ref{drude_1:eqn}),
on the contrary, hold for the positive-$U$ Hubbard as well.
\end{thebibliography}
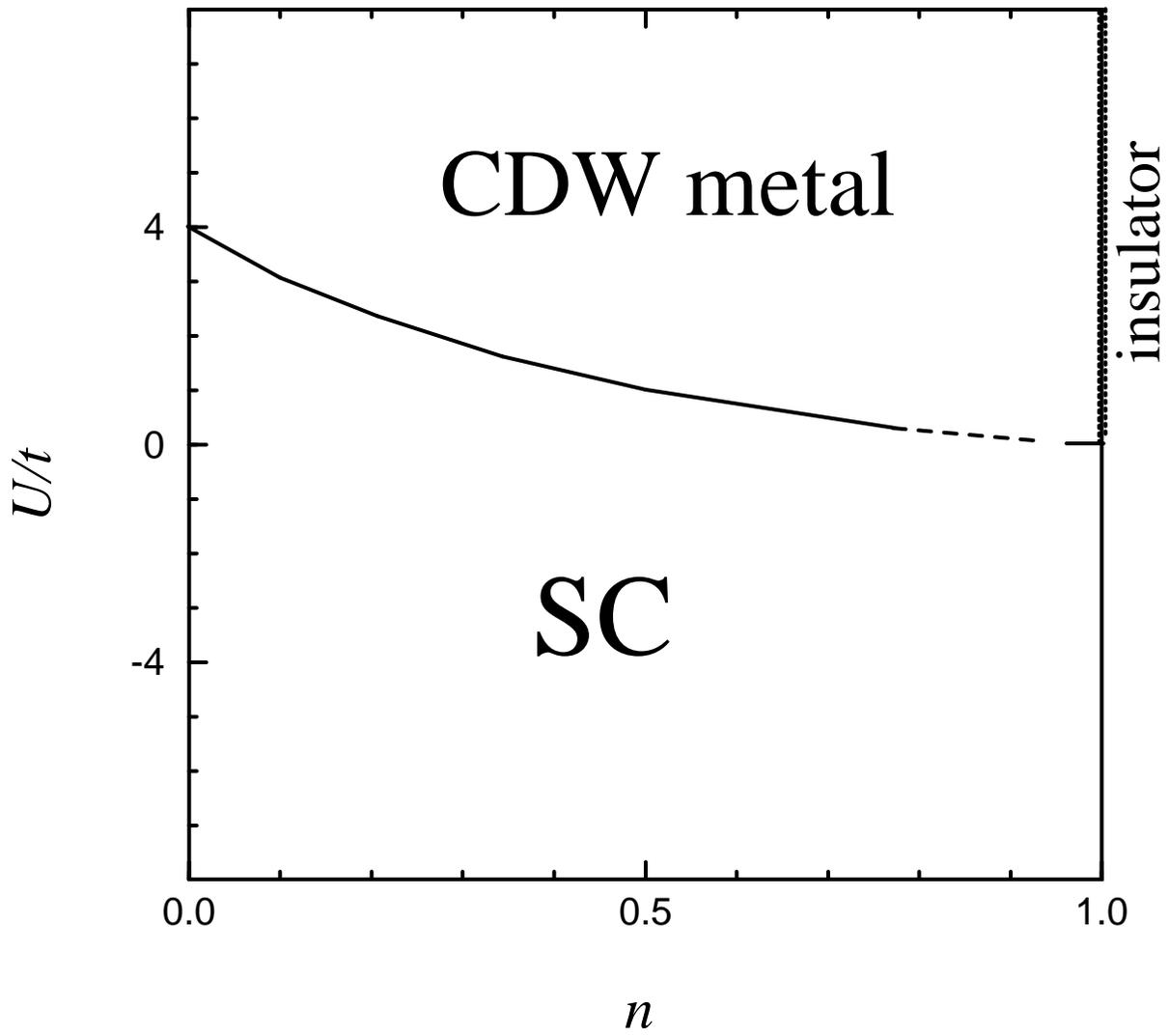
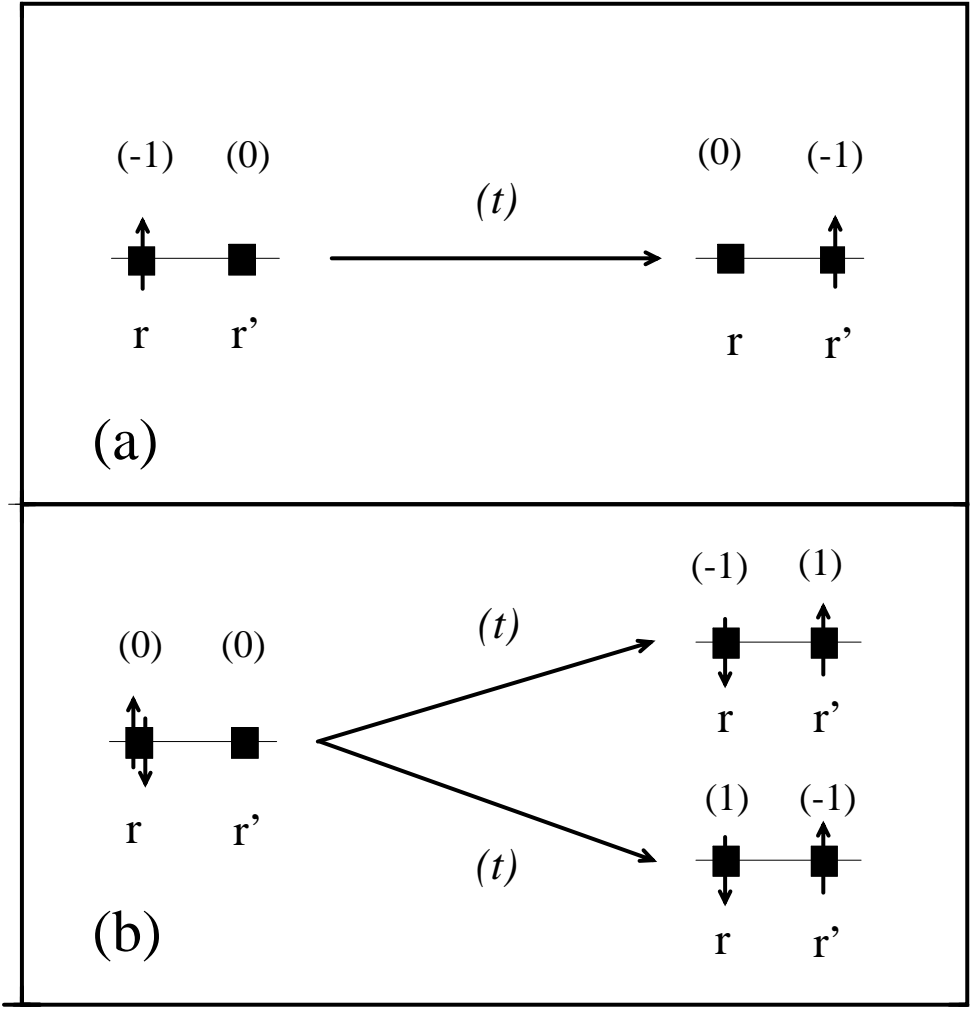
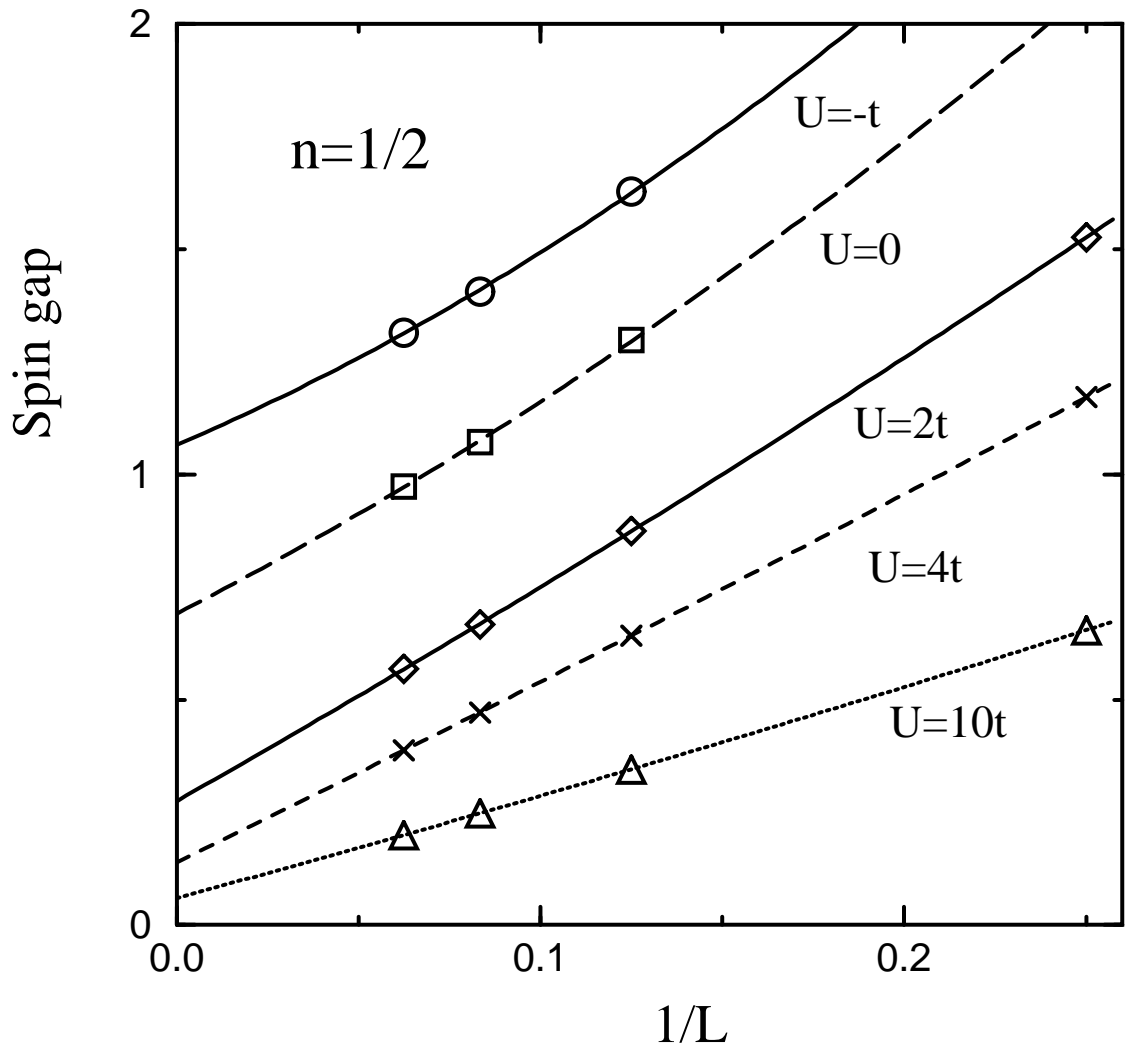
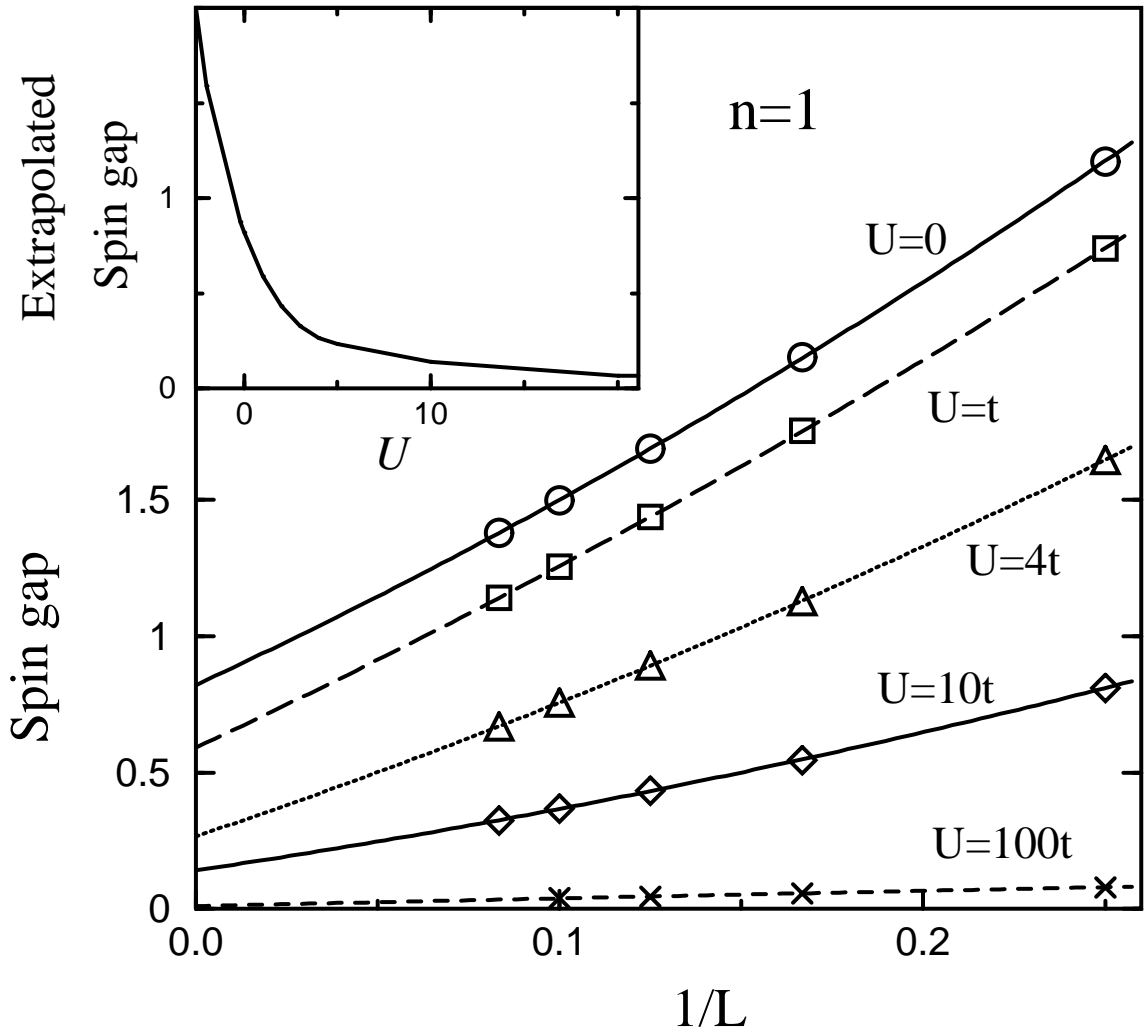
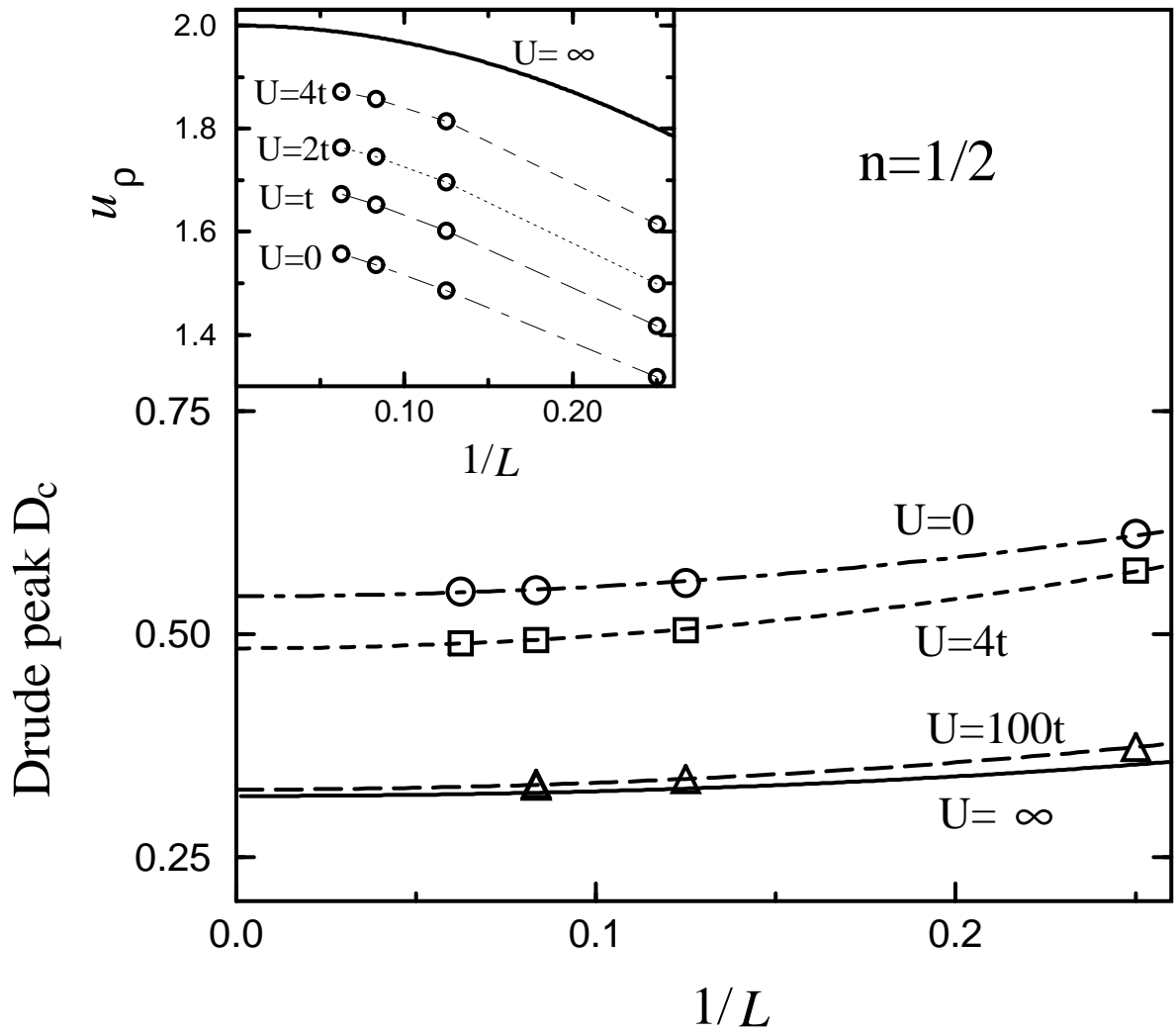
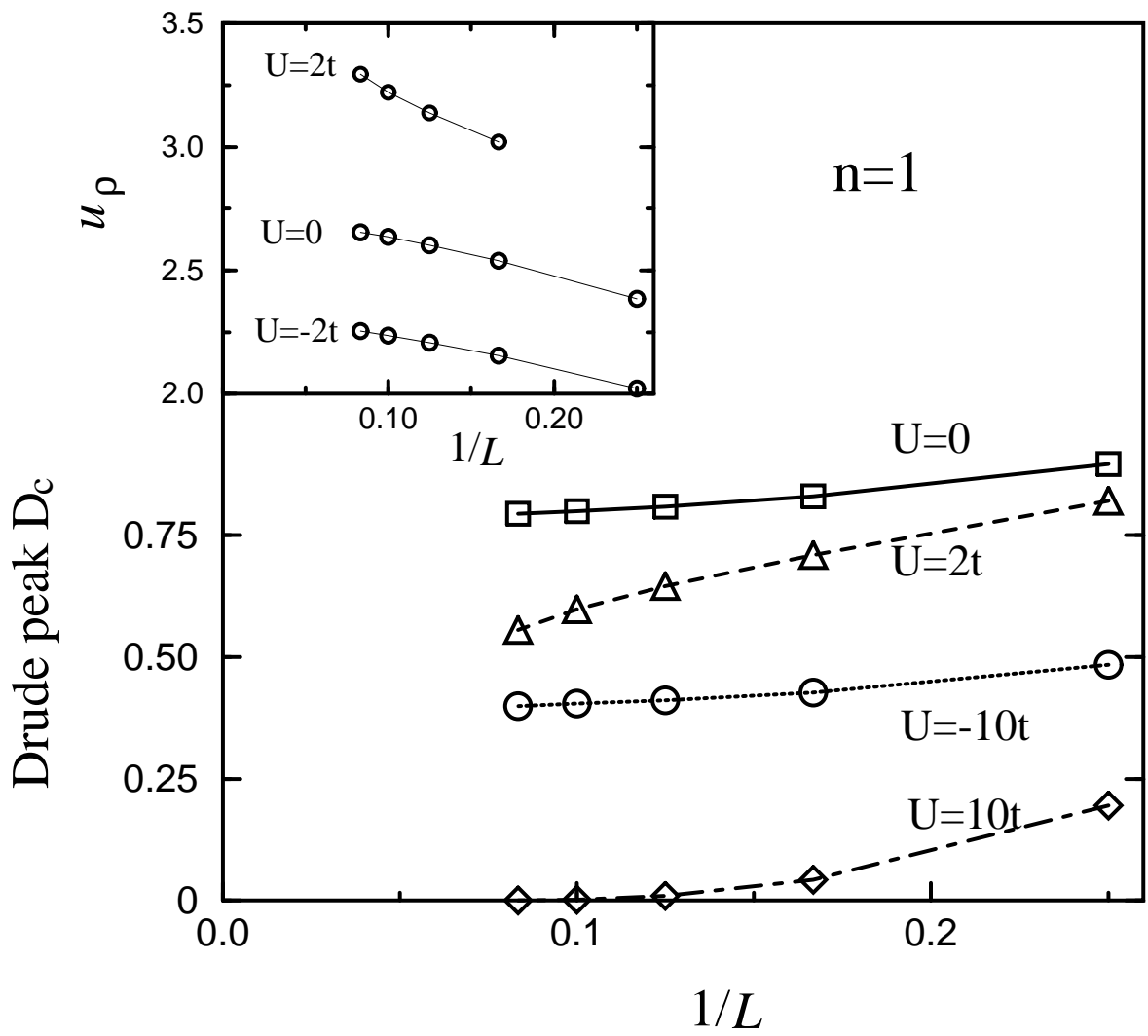
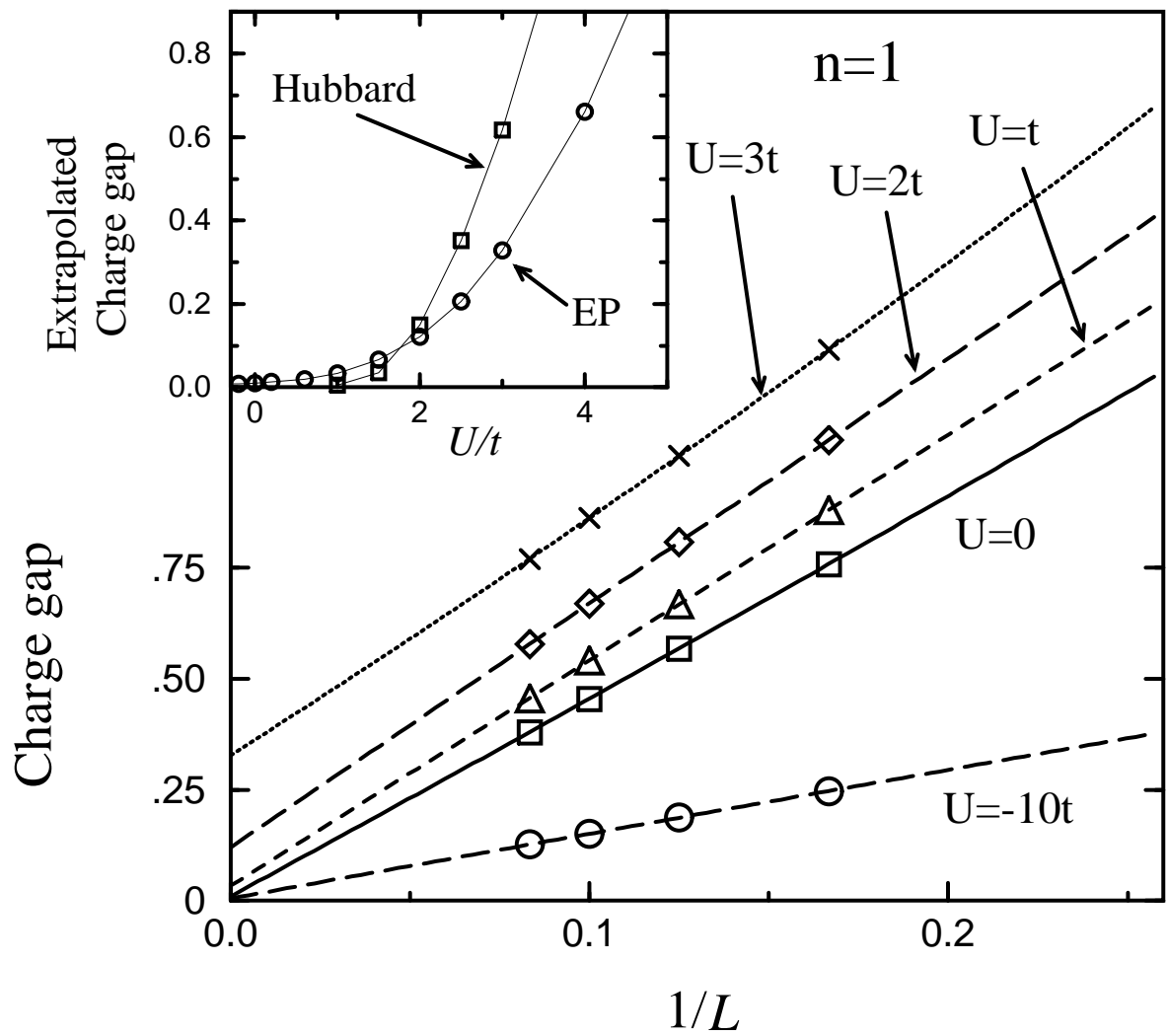
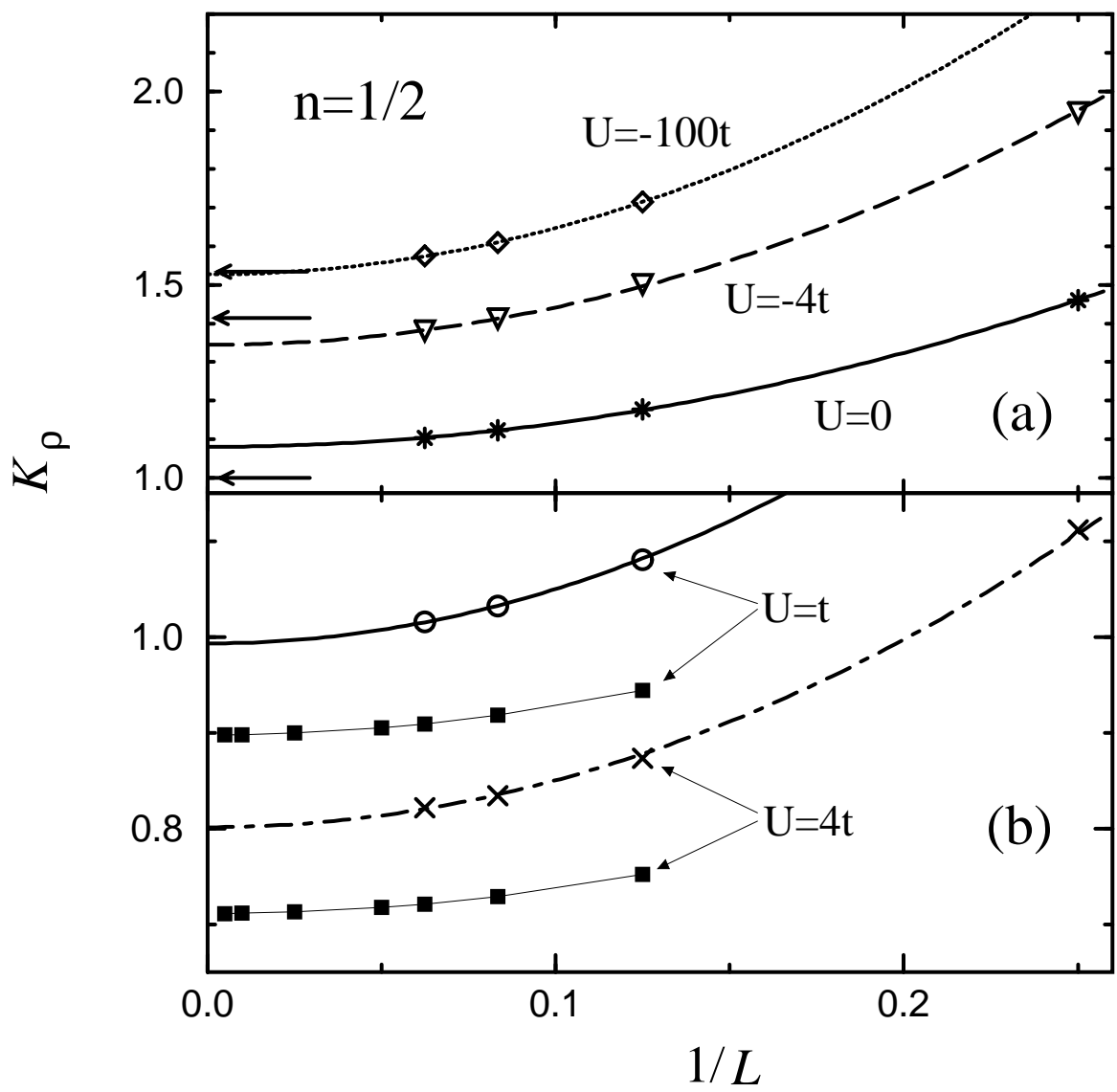
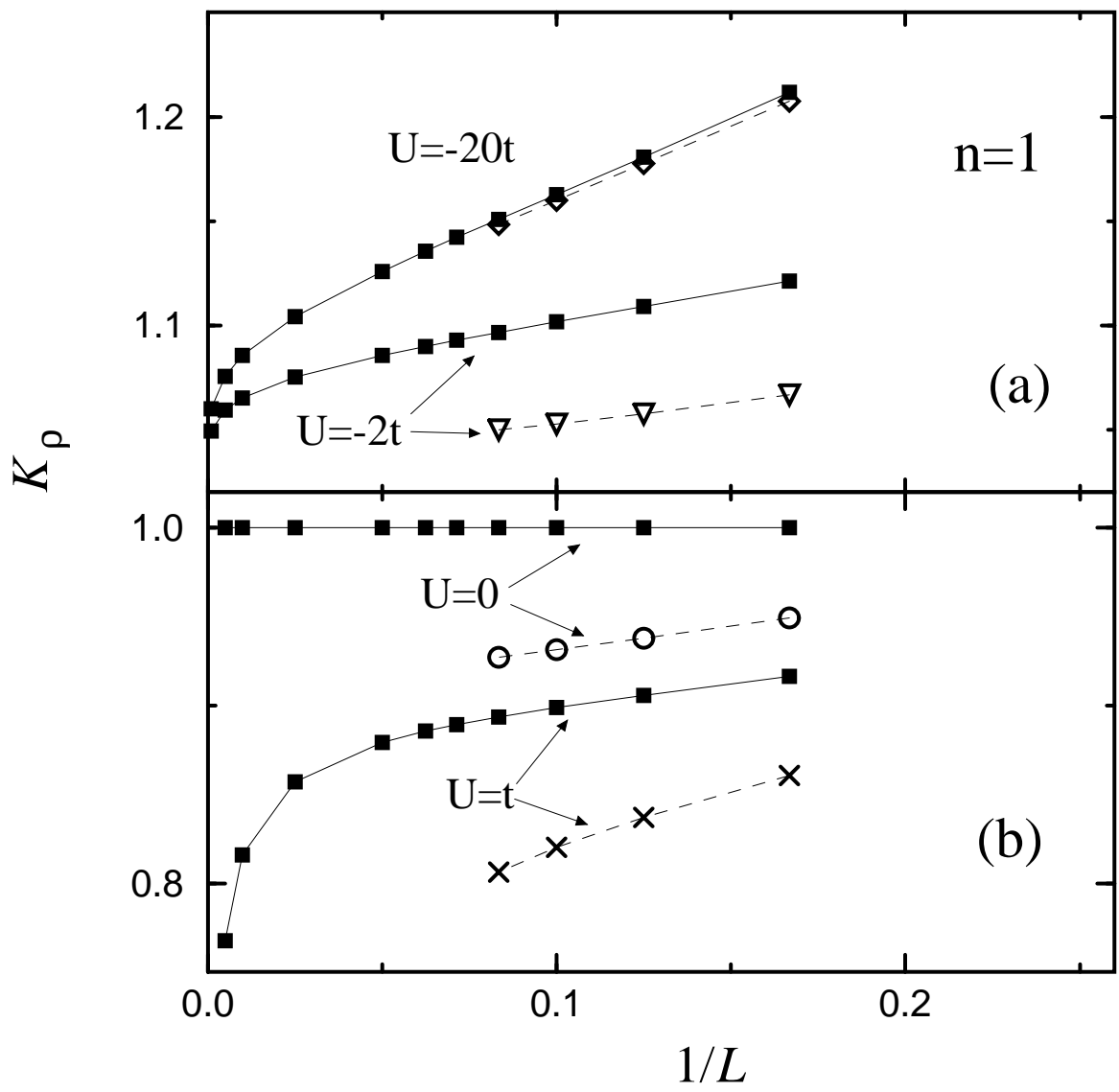
\end{document}